\documentclass[aps,prd,showpacs,showkeys,superscriptaddress,preprintnumbers,10pt,floatfix]{revtex4}

\usepackage{amsmath}
\usepackage{amssymb}  
\usepackage{graphicx}
\usepackage{aas_macros}
\usepackage[dvips]{color}
\usepackage{color}

\newcommand{\gsim}{\, \raisebox{-0.8ex}{$\stackrel{\textstyle >}{\sim}$ }}
\newcommand{\lsim}{\, \, \raisebox{-0.8ex}{$\stackrel{\textstyle <}{\sim}$ }}

\usepackage[normalem]{ulem}  

\newcommand{\bfq}{{\bf{q}}}

\newcommand{\bfr}{{\bf{r}}}
\newcommand{\ql}{q^{z}} 
\newcommand{\qp}{q_{\perp}} 

\newcommand{\Dslash}{D\!\!\!\!/}

\begin{document}
\title{Electronic screening and damping in magnetars}
\author{Rishi Sharma, Sanjay Reddy}
\affiliation{Theoretical Division, Los Alamos National Laboratory,
 Los Alamos, New Mexico 87545, USA}  
\begin{abstract}
We calculate the screening of the ion-ion potential due to electrons in the
presence of a large background magnetic field, at densities of relevance to
neutron star crusts. Using the standard approach to incorporate electron
screening through the one-loop polarization function, we show that the magnetic
field produces important corrections both at short and long distances.  In
extreme fields, realized in highly magnetized neutron stars called magnetars,
electrons occupy only the lowest Landau levels in the relatively low density
region of the crust. Here our results show that the screening length for Coulomb
interactions between ions can be smaller than the inter-ion spacing. More
interestingly, we find that the screening is anisotropic and the screened
potential between two static charges exhibits long range Friedel oscillations
parallel to the magnetic field. This long-range oscillatory behavior is likely
to affect the lattice structure of ions, and can possibly create rod-like
structures in the magnetar crusts. We also calculate the imaginary part of the
electron polarization function which determines the spectrum of electron-hole
excitations and plays a role in damping lattice phonon excitations. We
demonstrate that even for modest magnetic fields this damping is highly
anisotropic and will likely lead to anisotropic phonon heat transport in the
outer neutron star crust.   \end{abstract}
\keywords{neutron stars, magnetic fields, Friedel oscillations}
\pacs{97.60.Jd, 63.20.kd, 71.70.Di}
\maketitle

\section{Introduction}
Highly magnetized neutron stars called magnetars that feature extreme magnetic
fields ($B$) --- as large as $10^{15}$G on the surface and perhaps even larger
fields inside --- have been detected in recent years.  Currently there are 13
magnetar candidates which are classified as anomalous x-ray pulsars (AXPs) or
soft gamma repeaters (SGRs)
(http://www.physics.mcgill.ca/~pulsar/magnetar/main.html). Such large 
magnetic fields  can strongly influence the state of matter in
neutron stars, especially in the lower density region were the characteristic
energy scales associated with the matter fields is small compared to the
magnetic ``perturbation''. Several decades ago, it was realized that even modest
magnetic fields could influence the structure of atoms in the atmospheres of
magnetized neutron stars and affect the spectral features of the emitted
radiation \cite{Ruderman:1974} .  At the surface, where electrons are localized
around nuclei and the system consists of neutral atoms, large magnetic fields
distort the atomic structure resulting in rod-like or cigar shapes elongated
parallel to $B$. (For a recent review see \cite{Ruder:1994}.) It was realized
quite early that these cigar shaped atoms would bind with each other  along the
magnetic field direction forming polymer chains, and that these chains would
interact in the perpendicular plane, resulting in a condensed phase
\cite{Ruderman:1971}.  More recently the physics of these condensed phases for
hydrogen has been analyzed and its implications for neutron atmospheres is
discussed in Ref.  \cite{Lai:1997}. Here we will consider the effects of the
magnetic fields at higher density, in the crusts of magnetars where the
electrons are not localized in atomic states but rather form a degenerate Fermi
system. This will be important in determining the structural and transport
properties of the magnetar crust. Our finding will have implication for crust
oscillations (excited in explosive phenomena such as giant flares) and for heat
and electrical transport which are important ingredients in the study of
magnetic and thermal evolution of magnetars \cite{Geppert:2004,Aguilera:2007}. 

In this work we will typically restrict the study to strong fields where only a
few Landau levels are occupied and this will necessarily restrict us to
relatively low density region in the outer crust of the neutron star.  ($\rho
\lsim 10^{10}$ g/cm$^3$ for $B$ $\lsim$ 10$^{15}$ G.) The key microscopic quantity
required to characterize the response in our approach is the
polarization function, which is the Fourier transform of the density-density
correlation function. The expression for the polarization function for a
magnetized electron gas is calculated and is found in agreement with previous
work \cite{Danielsson:1995}.  Our new findings in this work are related to
identifying and understanding the implications of the electronic response in
strong magnetic fields for neutron star structure and transport. The real part
of the electron polarization function is related to the screening of the ion-ion
potential in the crust. In an unmagnetized crust, screening is dominated by the 
usual Debye screening. In neutron stars where the typical magnetic field is small ($B$
$\lsim10^{11}$ G),  the electron screening length is larger than the inter-ion
distance and this implies that the effects of electron screening are usually
unimportant in determining the crystal structure and thermodynamics. In
contrast, we find that for large fields when only one or a few Landau levels
are filled, the screening length can be either much smaller or much larger than
in the unmagnetized case depending on the density  and magnetic field. At
relatively low density the screening is enhanced resulting in a screening length
that is generically smaller than the inter-ion distance. Our finding of enhanced
screening for large $B$ is not new, this was already noted in early work which
explored some its implications for nuclear fusion reactions
\cite{Shalybkov:1987}. The qualitatively new finding of our work is the
realization that the sharp Fermi surface in the electron momentum distribution
along the $B$ field will give rise to long-range oscillatory behavior called
Friedel oscillations in the ion-ion potential --- a well studied feature in the
condensed matter context for $B=0$. The period of this spatial oscillation
$\lambda_F = \pi/(k_{f}^z)$, where $k_{f}^z$ the Fermi momentum of the electrons
in the $z$-direction, where we have assumed that $B$ is parallel to $z$. This
potential falls off slowly as $1/z$ on a scale of several inter-ion distances
and consequently can be very important in determining the phase structure of
the crust. We speculate that this will lead to the formation of an anisotropic
crystal with rod-like structures akin to the polymer chains expected in the
atomic regime.    

The imaginary part of the polarization is related to dissipation and gives rise
to the usual Landau damping observed in Fermi systems.  In the crust of a
neutron star, excitations such as lattice phonons couple to the electrons, and
their mean free paths are largely determined by the efficiency of the electronic
Landau damping where the phonon decays by producing an electron-hole pair. We
show that this decay rate has a non-trivial dependence on the angle between
their propagation direction and the magnetic field because the electron-hole
excitation for large $B$ is highly anisotropic. This anisotropy leads to an
anisotropy in heat transport properties of lattice phonons. We find that the
lattice phonons can be scattered more easily if they are moving perpendicular to
the field, compared to when they are moving parallel to it. 

The paper is organized as follows. In Section~\ref{Section:PolarizationMagnetic}
we derive the polarization function of electrons in the presence of a magnetic
field. In the Sections~\ref{Section:Screening} and \ref{Section:HeatTransport},
we apply the result to two physical problems. In Section~\ref{Section:Screening}
we present numerical results for Friedel oscillations. In
Section~\ref{Section:HeatTransport} we discuss the effect of anisotropic
polarization on the decay rates of the lattice phonons as a function of the
angle between the propagation direction and the magnetic field.

 We discuss the implications for the structure of the magnetar crust in Section~\ref{Section:Conclusions}. Details of calculations of the expression for
the screened potential are given in Appendix~\ref{Section:LLL dominates} and 
Appendix~\ref{Section:Friedel derivation}.
\section{Polarization of electrons in a magnetic field.\label{Section:PolarizationMagnetic}}

The Lagrangian for the electron gas in an external electromagnetic field and at finite chemical potential $\mu$ is given by,
\begin{equation}
L = \int d^3x \bar{\psi}(i\Dslash-m_e + \mu\gamma^0)\psi\;,
\end{equation}
where $D_\mu = \partial_\mu - ie A_\mu$ and $m_e$ is the electron mass. We consider a time independent, uniform magnetic 
field $B$ in the ${z}$ direction and choose a gauge such that the only non-zero component of the gauge
field $A$ is $A^x = -A_x = By$. The spectrum of electron energy levels in an external magnetic field is well known 
and electrons occupy Landau levels with energies given by 
\begin{equation} 
E_{m} = \sqrt{(k^z_{m})^2+2meB+m_e^2}\,, 
\end{equation} 
where $m$ is the quantum number associated with the Landau level and $k^z_m$ is the momentum in the $z$ direction. 
The number density of electrons is
\begin{equation}
n_e =\frac{eB}{(2\pi)^2}(\int_{-\infty}^{\infty} dk^z f_0 + 
 2\sum_{m>0}\int_{-\infty}^{\infty} dk^z f_m)
 \label{eq:ne}
\end{equation}
where $f_{m} = (\exp(\frac{E_{m}-\mu}{T})+1)^{-1}$ is the Fermi distribution function. At $T=0$, we obtain
\begin{equation}
n_e =\frac{eB}{2\pi^2}(\sqrt{\mu^2-m_e^2} + 
 2\sum_{m>0}\theta(\mu^2-m_e^2-2meB)\sqrt{\mu^2-m_e^2-2meB})\;.
 \label{eq:neT0}
\end{equation}
We will often restrict to the low-density or large magnetic field limit, where only the lowest Landau level is occupied. This occurs for $2eB>\mu^2-m_e^2$.

The density-density correlation function is defined as 
\begin{equation}
\chi_B(x^\mu,y^\mu) = \frac{1}{i}\langle T\{n(x)n(y)\}\rangle\label{eq:chiB}\,, 
\end{equation}
where $n(x)= \psi^\dagger(x) \psi(x)$ is the density operator. The Lagrangian is time independent for a static external field and even though the Lagrangian looks position
dependent in a particular gauge, the system itself is translationally invariant. (The change in the
Hamiltonian due to a translation in ${x}$ can be undone by a gauge transformation.) Therefore we can write the correlation function in momentum space as,
\begin{equation}
\chi_B(x^\mu,y^\mu) = \int \frac{d\omega}{2\pi} \frac{1}{V} \sum_q e^{iq_\mu (x-y)^\mu}
\Pi_B(\omega,{\bf{q}})\label{eq:piB}\;,
\end{equation}
where $q^\mu=(\omega,\bfq)=(\omega,\qp,\ql)$.

To evaluate $\Pi_B$, we insert a complete set of many-body eigenstates between the two density
operators in Eq.~\ref{eq:chiB}. We ignore the anti-particle contribution to the polarization function because 
they are well below the Fermi level and retain only particle-hole contributions. In this approximation
we can write the polarization as,
\begin{equation}
\begin{split}
\Pi_B(q^\mu) 
=& \frac{eB}{2\pi} \sum_{m,n} \int_{-\infty}^\infty \frac{dk^z_m}{2\pi}f_m (1-f_n)
 W(E_m,E_n,m,n,k^z_m,k^z_n,\qp)\\
 &\bigl[\frac{1}{\omega-E_n+E_m+i\epsilon} 
-\frac{1}{\omega+E_n-E_m-i\epsilon} \bigr]|_{k^z_n=k^z_m+\ql}\label{eq:PiB simplified}\;,
\end{split}
\end{equation}
where, $m$ refers to the (initial) energy eigenstate with energy $E_{m} = \sqrt{(k^z_{m})^2+2meB+m_e^2}$ and $n$ to the (final) state with energy $E_{n} = \sqrt{(k^z_{m}+q^z)^2+2neB+m_e^2}$. 
 The function $W$ contains the matrix elements of the density operator between different electron eigenstates and is given by ,
\begin{equation}
\begin{split}
W = W(E_m,E_n,m,n,k^z_m,k^z_n,\qp) =& \frac{1}{2E_nE_m}\bigl[(E_mE_n+k^z_mk^z_n+m_e^2)H_{m-1,n-1}(\qp)H^*_{m-1,n-1}(\qp)\\
 &+(\sqrt{2meB}\sqrt{2neB})H_{m-1,n-1}(\qp)H^*_{m,n}(\qp)\\
 &+(\sqrt{2meB}\sqrt{2neB})H_{m,n}(\qp)H^*_{m-1,n-1}(\qp)\\
 &+(E_mE_n+k^z_mk^z_n+m_e^2)H_{m,n}(\qp)H^*_{m,n}(\qp) \bigr]\label{eq:W defined}\;,
\end{split}
\end{equation}
where,
\begin{equation}
H(\qp)=\left\{\begin{array}{cc}
m\geq n &\bigl[ e^{-\qp^2l^2/4}\sqrt{\frac{n!}{m!}} \Bigl(\frac{(-q^x-iq^y)l}{\sqrt{2}}\Bigr)^{m-n}
{{L}}_n^{m-n}(\qp^2l^2/2)\bigr]\\
m\leq n &\bigl[ e^{-\qp^2l^2/4}\sqrt{\frac{m!}{n!}} \Bigl(\frac{(q^x+iq^y)l}{\sqrt{2}}\Bigr)^{n-m}
{{L}}_m^{n-m}(\qp^2l^2/2) \bigr]
\end{array}\right.\label{eq:H defined}\;\;.
\end{equation}
Here, ${{L}}_a^b(x)$ are the associated Laguerre polynomials, and $l=1/\sqrt{eB}$ is the magnetic length. 
For $m=0$ or $n=0$, $H_{m-1,n-1}(X)=0$.


The imaginary and real parts of the response function can be separated by following the pole
prescription,
\begin{equation}
\frac{1}{\omega - \Delta E \pm i \epsilon} \
  = {\rm{P}}\bigl(\frac{1}{\omega - \Delta E  }\bigr) \mp i\pi \delta(\omega - \Delta E )\;.
\end{equation}
This gives the imaginary part,
\begin{equation}
\begin{split}
\Im m[\Pi_B(q^\mu)] =& \frac{eB}{2\pi} (-\pi)\sum_{m,n} \int_{-\infty}^\infty \frac{dk^z_m}{2\pi}
 f_m (1-f_n)W(E_m,E_n,m,n,k^z_m,k^z_n,\qp)\\
& \bigl[\delta({\omega-E_n+E_m}) 
-\delta({\omega+E_n-E_m}) \bigr]|_{k^z_n=k^z_m+\ql}\label{eq:ImPi}\;,
\end{split}
\end{equation}
We want to consider situations where the temperature is much less than the chemical potential $\mu$.
In this case, it is appropriate to approximate $f_m$ by $\theta(\mu-E_m)$ and $1-f_n$ by
$\theta(E_n-\mu)$.


The real part of $\Pi_B$ is given by,
\begin{equation}
\begin{split}
\Re e[\Pi_B(q^\mu)] 
=& \frac{eB}{2\pi} \sum_{m,n}\int_{-\infty}^\infty \frac{dk^z_m}{2\pi}
 f_m W(E_m,E_n,m,n,k^z_m,k^z_n,\qp)\\
 &{\rm{P}}\bigl[\frac{1}{\omega-E_n+E_m} 
-\frac{1}{\omega+E_n-E_m} \bigr]|_{k^z_n=k^z_m+\ql}\label{eq:RePi}\;,
\end{split}
\end{equation}
In particular, to calculate screening, we will be interested in the static response, 
$\omega=0$, which is given by,
\begin{equation}
\begin{split}
\Re e[\Pi_B(0,\bfq)] =
& \frac{eB}{\pi}  \sum_{m,n}\int_{-\infty}^\infty \frac{dk^z_m}{2\pi}
 f_m W(E_m,E_n,m,n,k^z_m,k^z_n,\qp)\\
 &{\rm{P}}\bigl[\frac{1}{E_m-E_n} \bigr]|_{k^z_n=k^z_m+\ql}\label{eq:RePiw0}\;.
\end{split}
\end{equation}
In the limit $T=0$,
\begin{equation}
\begin{split}
\Re e[\Pi_B(0,\bfq)] =& \frac{eB}{2\pi} 2 \sum_{m,n}\int_{-\infty}^\infty \frac{dk^z_m}{2\pi}
 W(E_m,E_n,m,n,k^z_m,k^z_n,\qp)\\
 &\theta(\mu-\sqrt{(k^z)^2+m_e^2+2meB}){\rm{P}}\bigl[\frac{1}{E_m-E_n}
 \bigr]|_{k^z_n=k^z_m+\ql}\label{eq:RePiw0T0}\;.
\end{split}
\end{equation}
We will now use these results for calculating two important physical processes of relevance to magnetars: (i) the screening of electric charge
by the electron gas and the resulting modification of the ion-ion potential in the crust (Section~\ref{Section:Screening}), and (ii) the anisotropic spectrum of electron-hole excitations (Landau damping) and its consequences for the phonon mean free paths in the outer crust (Section~\ref{Section:HeatTransport}).

\section{Screening in an electron gas\label{Section:Screening}}
The screened potential between two ions with charges 
$Z_1$ and $Z_2$ separated by a displacement $\bfr$ in an electron gas is given by 
\begin{eqnarray} 
V_{\rm mod}(\bfr) &=& \frac{Z_1 Z_2~e^2}{4 \pi r}g(\bfr)\;,  \nonumber \\
{\rm where} \quad g(\bfr) &=& {4\pi r}\int\;\frac{d^3q}{(2\pi)^3} \;\frac{e^{i\bfq\cdot\bfr}}{q^2+F(\bfq)}
\label{eq:vmod}
\end{eqnarray}
gives the modification of the bare interaction due to the polarizability of the
electron gas. The function $F(\bfq)=-e^2\Re e[\Pi_{00}(\omega=0,\bfq)]$ is
related to the static polarization function of the electron gas,
$\Pi_{00}(0,\bfq)$. In evaluating the integral in Eq.~\ref{eq:vmod}, typically
it is assumed that for distances $r \gsim 1/k_f$, where $k_f$  is the Fermi
momentum of the electrons, we can replace $\Pi_{00}(0,\bfq)$ by
$\Pi_{00}(0,{\bf{0}})$.  In doing so one obtains the usual Debye screened
potential in coordinate space
\begin{equation} 
V_{\rm mod}(r) =  \frac{Z_1 Z_2~e^2}{4 \pi r}~   \exp(-r/\lambda_D)\,,
\end{equation} 
or $g(r) = \exp(-r/\lambda_D)$ where $ \lambda_D=1/\sqrt{F({\bf{0}})}$ is the Debye screening length, and is simply related to the 
Debye mass $m_D$, by $\lambda_D=1/m_D$. F({\bf{0}}) can be calculated using the well known compressibility sum-rule 
\cite{FetterWalecka:1971} and is given by 
\begin{equation} 
F({\bf{0}}) = e^2~\frac{\partial n_e}{\partial \mu_e}\,, 
\label{eq:fzero}
\end{equation} 
where $n_e$ is the equilibrium density and $\mu_e$ is the chemical potential and $e^2=4\pi/137$.

  For a free electron gas $F({\bf{0}}) =  e^2k_f \mu/\pi^2$ with
$\mu=(k_f^2+m_e^2)^{1/2}$.  The ion density $n_I=3/(4\pi a^3)$ --- where $a$ is
the average inter-particle separation (for a simple cubic crystal
$a$ is related to the lattice spacing $L$ by $(4\pi/3)^{1/3}a=L$) --- and the electron
density are related by electric charge neutrality which requires that
$n_e=Zn_I$, where $Z$ is the charge of each ion. Consequently, the electron
Fermi momentum $k_f$ and the average inter-ion distance $a$ are related by
$k_f=(3\pi^2 Z n_I)^{1/3} =5.66~(Z_{26})^{1/3}~a^{-1}$ where $Z_{26}=Z/26$.
Further, from Eq.~\ref{eq:fzero} we have $\lambda_D= 10.37\sqrt{v_f}/k_f$
where $v_f=k_f/\mu$. The ratio of the screening length to the average
inter-ion distance is therefore given by 
\begin{equation}
\frac{\lambda_D}{a} \simeq \frac{1.83}{(Z_{26})^{1/3}}\sqrt{v_f}\label{eq:lDvsa} \,, 
\end{equation}
implying that in the relativistic regime, where $k_f \gg m_e$ and $v_f \simeq
1$, $\lambda_D\simeq 1.83~a$. Since the interaction between the nearest ionic
neighbors is essentially unscreened in the relativistic electron gas, screening
is not considered to be important for determining the structure of the neutron
star crust. This situation is changed if the electrons are non-relativistic. For
$Z_{26}=1$, $\lambda_D<a$ for  $v_f<0.3$. However at lower densities, where
$v_f \ll 1$, orbital effects from the localization of electrons may be important 
and a simple screening approximation is not valid. 

\begin{figure}[t]
   \centering
   \includegraphics[width=4in]{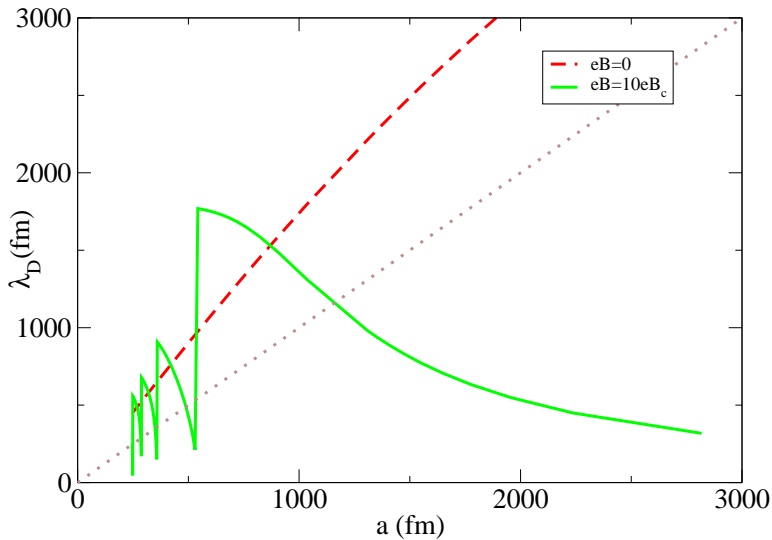} 
   \caption{(color online) The screening length $\lambda_D$ versus the inter-particle separation $a$
   for $Z=26$ and $eB=10eB_c$. The dotted line corresponds to $\lambda_D=a$. For
   $a\gtrsim512$(fm), the lowest Landau level is occupied, and for large regions
   $\lambda_D<a$ . As $a$ decreases, more levels are occupied.}  
   \label{fig:lambdavsa}
\end{figure}

Extending the Debye approximation to an electron gas in a magnetic field,
setting $F_B(\bfq)$ to its long-wavelength limit $F_B(\bfq={\bf{0}})$, we can
deduce how the Debye screening is modified by the magnetic field. When only one
Landau level is occupied, we find 
\begin{eqnarray}
\lambda_D(2eB>{\mu_e^2-m_e^2}) &=& \frac{\pi \sqrt{2 }}{e\sqrt{eB}}~\sqrt{v_{f}^z} \label{eq:lDvsaB}\,, 
\end{eqnarray}
which can be smaller or larger than $a$. Hence the screening length between ions
relative to their separation can tuned arbitrarily by introducing a large
magnetic field. To illustrate this we plot $\lambda_D$ versus $a$ for
$eB=10eB_c$ in Fig.~\ref{fig:lambdavsa}. In the low density regime where only
one Landau level is filled, $\lambda_D < a $ for $a>1164$(fm). With increasing
density (smaller $a$), $\lambda_D$ becomes larger than $a$ and continues to
oscillate around the $B=0$ value as more Landau levels are occupied. The
amplitude of these oscillations rapidly decreases with $n$ where $n$ is the
number of Landau levels filled. As noted earlier, for $B=0$ the screening length
is always larger than $a$ when electrons are relativistic.  

At low temperatures the Fermi surface of the electrons is sharp and this gives
rise to non-analyticities in $F(\bfq)$. Therefore it is not possible to
approximate $F(\bfq)$ in Eq.~\ref{eq:vmod} by $F({\bf{0}})$ even for $r\gg
1/k_f$. The screened potential in position space depends on the polarization
function for values of $\bfq$ other than $\bfq={\bf{0}}$, and has qualitative
new features compared to the Debye screened result. This phenomenon is well known for the free
electron gas ($B=0$)~\cite{FetterWalecka:1971}.

Kapusta and Toimela~\cite{KapustaToimela:1988} employed the fully relativistic polarization function for electrons at one-loop order to calculate the most general form of the screened potential
in a free electron gas. The central result of their calculation is that $F(\bfq)$ has branch cuts along $q=|\bfq|=\pm2k_f+i\eta$ for $\eta>0$.
Consequently at large distances, where $r \gg \lambda_D$ the screened potential exhibits characteristic oscillations called Friedel oscillations. It is well established~\cite{FetterWalecka:1971} that in the non-relativistic limit the screening function is given by
\begin{equation}
g_{\rm Friedel(NR)}(r\gg \lambda_D)= \frac{4\xi^2}{(4+\xi)^2}\frac{\cos{(2 k_f r)}}{( r/\lambda_D)^2} \,, 
\end{equation} 
where $\xi=m_{D}^2/2 k_f^2= (e^2 \mu)/(\pi^2 k_f)$. At very low density, where $\xi
\gsim 1$  Friedel oscillations are non-negligible and have been well studied in condensed matter
physics context~\cite{FetterWalecka:1971}. The potential between two ions for $r\lesssim \lambda_D$ is dominated by Debye screening.  In the region $r \simeq \lambda_D$ a simple analytic
form does not exist and the screening function needs to be calculated numerically.

It is also shown in Ref.~\cite{KapustaToimela:1988} that in the extreme relativistic limit, when $m_e\rightarrow 0$, Friedel oscillations are modified. At large distances the screening function
is of the form
\begin{equation}
g_{\rm Friedel(R)}= (2\xi)^{3/2}\frac{\sin{(2 k_f r)}}{( r/\lambda_D)^3} \,.  
\end{equation} 
Unlike in the non-relativistic regime, here $ \xi = 0.0047 \ll 1$, and amplitude of these oscillations
is greatly suppressed. Furthermore they decay as $1/r^3$ instead of $1/r^2$ in the non-relativistic
limit. Consequently, Friedel oscillations can be ignored in the relativistic regime.

In the presence of a magnetic field, the screening function $F_B(\bfq)$ is no
longer rotationally symmetric and we expect the screened potential to be
anisotropic. We now proceed to calculate $V(\bfr)$ in this case. The calculation
of the polarization tensor when the densities are high and several Landau levels
are occupied, is technically complicated. On the other hand, in this limit one
expects the results to
be very similar to the free electron gas. Therefore we restrict ourselves to the large field limit,
when only the lowest Landau level is filled. In cylindrical variables we write $F_B(\bfq)$ as a function of $\ql=q \cos{\theta}$ and $\qp=q \sin{\theta}$, where $\theta$ is the angle between the ${\bf{q}}$ and the magnetic field ${\bf{B}}$.  Note that Debye screening --- which depends only on $F_B({\bf{0}})$ --- is completely isotropic.

In cylindrical spatial coordinates the potential between two ions 
\begin{eqnarray} V_{\rm mod}(\rho,z) &=&
\frac{Z_1Z_2e^2}{4\pi r}g(\rho,z)  \;, \nonumber \\
{\rm where} \quad g(\rho,z) &=&
\frac{r}{\pi}\int_0^\infty d\qp \qp J_0(\qp\rho)\int_{-\infty}^\infty d\ql
\frac{e^{i\ql z}}{q^2+F_B(\qp,\ql)} \,, \label{eq:vmodmag}
\end{eqnarray}
with $r=\sqrt{\rho^2 + z^2}$, $F_B(q_\perp,q^z)=-e^2 \Pi_B(0,{\bf{q}})$, and
$\Pi_B(0,{\bf{q}})$ is the static polarizability of the electron gas in the presence of a
magnetic field.

Restricting to $m=0$ in the sum in Eq.~\ref{eq:RePiw0T0}, we have,
\begin{equation}
\begin{split}
F_{B}(q_\perp,q^z) &= e^2\frac{eB}{2\pi}\sum_{n}\int_{-\infty}^\infty\frac{dk^z}{2\pi}
2\theta(\mu-\sqrt{(k^z)^2+m_e^2})\frac{W(E_0,E_n,m,n,k^z_0,k^z_n,q^{z})}{E_n-E_0}\,. \\
\label{eq:Fb}
\end{split}
\end{equation}
The sum over $n$ in Eq.~\ref{eq:Fb} runs over all
non-negative integers, but the most important contribution for $eB\gg\mu^2$ and 
$eB\gg q^2$ is the $n=0$ term. (This is shown in Appendix~\ref{Section:LLL dominates}.) 
Therefore for simplicity, we drop the $n>0$ terms, which allows us to calculate analytic 
expressions for $F_B(\bfq)$, and the screened Coulomb interaction in certain limits. With these approximations,
\begin{equation}
\begin{split}
F_{B}(\qp,\ql) \sim (\frac{e}{\pi})^2&\bigl(\frac{eB}{2}\bigr)e^{-\qp^2/(2eB)} \int_{-k_f^z}^{k_f^z}
d k^z\frac{1}{\sqrt{(k^z+\ql)^2+m_e^2}-\sqrt{(k^z)^2+m_e^2}}\\
&\frac{((k^z+\ql)k^z+m_e^2+\sqrt{(k^z)^2+m_e^2}\sqrt{(k^z+\ql)^2+m_e^2})}
{2\sqrt{(k^z)^2+m_e^2}\sqrt{(k^z+\ql)^2+m_e^2}}\,, 
\end{split}
\end{equation} 
where $k_f^z=\sqrt{\mu^2-m_e^2}$ is the Fermi momentum of the electrons in the $z$ direction for the $m=0$ level. If higher levels are occupied, each will have a different Fermi momentum. 

\begin{figure}[ht]
   \centering
   \includegraphics[width=5in]{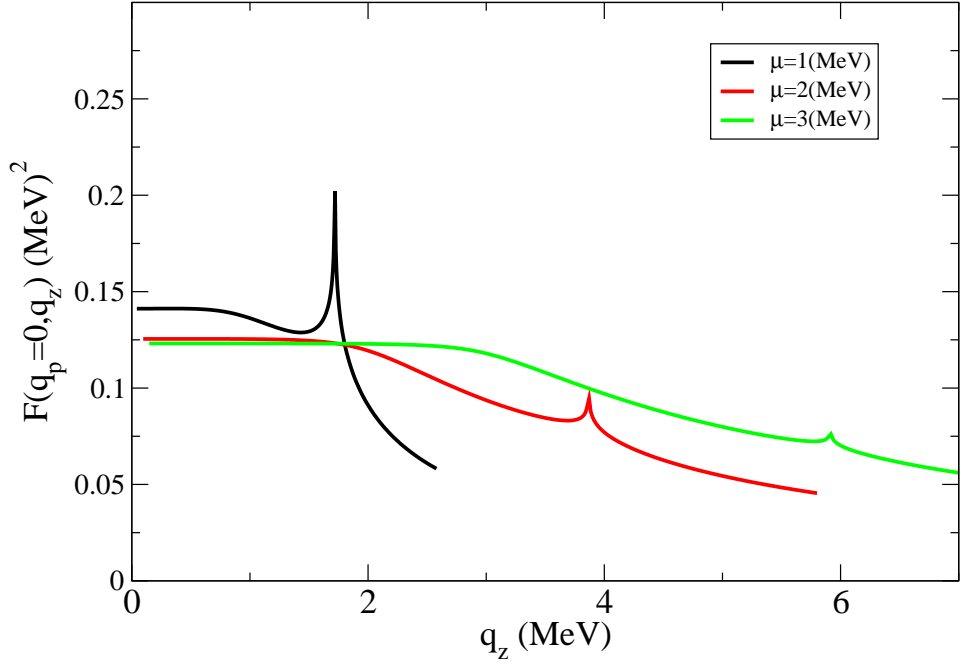} 
   \caption{(color online) $F_B$ as a function of $q^z$ for $q_\perp=0$, $eB=100eB_c$, and three values of $\mu$. As the
   electrons become more relativistic ($\mu$ increases), the kink at $2k_f^z$ becomes less sharp. The 
   shoulder at $k_f^z$ becomes non analytic in the ultra-relativistic limit.}  
   \label{fig:FBqz}
\end{figure}

Now we consider two limiting cases. First, in the non-relativistic regime where
$m_e\gg k_f^z$, we obtain for large $B$,
\begin{equation}
\begin{split}
F_B(\qp,\ql)&\sim (\frac{e}{\pi})^2\bigl(\frac{eB}{2}\bigr)e^{-\qp^2/2eB} \int_{-k_f^z}^{k_f^z}
d k^z \bigl(\frac{m_e}{(k^z+\ql)^2-(k^z)^2}\bigr)\\
&=
 \bigl(\frac{e}{\pi}\bigr)^2\bigl(\frac{eB}{2}\bigr)e^{-\qp^2l^2/2}
 \log\bigl(\frac{2k_f^z+\ql}{2k_f^z-\ql}\bigr)\frac{m_e}{\ql}\label{eq:FB nonrel}\;.
\end{split}
\end{equation}
The Debye mass is given by 
\begin{equation}
m_D^2 = \lim_{\bf{q}\rightarrow 0} F_B(\qp,\ql) =
(\frac{e}{\pi})^2(\frac{eB}{2})(\frac{m_e}{k_f^z})  \label{eq:mDnrel}\,,
\end{equation}	
and since $n_e=eB/(2\pi^2)\sqrt{\mu^2-m_e^2}$, the above expression
(Eq.~\ref{eq:mDnrel}) is consistent with Eq.~\ref{eq:fzero}. The key feature of the expression for
$F_B(\qp,\ql)$ in Eq.~\ref{eq:FB nonrel} is its non-analytic behavior as a
function of $q^z$. In the complex plane, Eq.~\ref{eq:FB nonrel} has branch
cuts along $q^z=\pm 2k_f^z + i\eta$, $\eta>0$.  Restricting $q^z$ to the real
axis, one sees a kink at $q^z=\pm 2k_f^z$ (Fig.~\ref{fig:FBqz}). On the other
hand the expression is analytic in $q_{\perp}$.  Based on this we can expect
long range oscillations in $g(z)$ with wavelength $\pi/k_f^z$ in the $z$
direction, but no long range features in the $x-y$ plane.

In the relativistic regime, where $m_e\ll\mu$ we find that 
\begin{equation} 
\begin{split}
F_{B}(\qp,\ql) &\sim (\frac{e}{\pi})^2\bigl(\frac{eB}{2}\bigr)e^{-\qp^2/2eB}
\int_{-k_f^z}^{k_f^z}
d k^z \frac{(k^z+\ql)k^z+|k^z||k^z+\ql|}
{2|k^z||k^z+\ql|}\bigl(\frac{1}{-|k^z|+|k^z+\ql|}\bigr)\\
&= \left\{
\begin{array}{cc}
(\frac{e}{\pi})^2 \bigl(\frac{eB}{2}\bigr) e^{-\qp^2/2eB} & \qp < k_f^z  \\
(\frac{e}{\pi})^2 \bigl(\frac{eB}{2}\bigr) e^{-\qp^2/2eB}~\frac{k_f^z}{|\ql|} & \qp > k_f^z 
\end{array}
\right.\;.\label{eq:FB rel}
\end{split}
\end{equation}

The debye mass square is $m_D^2 = \lim_{\bf{q}\rightarrow 0} F_B(\qp,\ql) = 
(\frac{e}{\pi})^2(\frac{eB}{2})$. The number of electrons in these limits 
is $n_e=eB/(2\pi^2)\sqrt{\mu^2-m_e^2}\sim (eB/(2\pi^2))\mu$ and therefore $m_D^2=e^2
dn_e/d\mu_e=e^2 (eB/(2\pi^2))$ as expected.

From Eq.~\ref{eq:FB rel} we note that in the ultra-relativistic limit $F_{B}$
does not have any non-analytic behavior at $q^z=2k_f^z$. On the other hand $F_B$
is non-differentiable at $q^z=k_f^z$.  This behavior is different from that of a
free electron gas, where the non-analyticity occurs at $\pm 2k_f^z$ in both the
relativistic and non-relativistic limits. However, this is an artifact of the
ultra-relativistic limit, and for any finite $m_e$, $F_B$ is differentiable
at $q^z=k_f^z$. We plot $F_B(q^z)$  for three values of $\mu$ in
Fig.~\ref{fig:FBqz}.  We notice that for fixed $\mu$ there are two distinct
features in the plot versus $q^z$. First, there is shoulder at $k_f^z$ which
becomes sharper as $m_e/\mu$ decreases. Second, there is a kink at $2k_f^z$
which becomes less prominent as $m_e/\mu$ decreases.  In the ultra-relativistic
limit the kink disappears, and the shoulder becomes a non-differentiable point.
However, for any finite $m_e$, the fact that the kink becomes less prominent
suggests that the Friedel oscillations will become weaker as $m_e/\mu$
decreases. Hence, we will now focus on the non-relativistic limit. 

In the non-relativistic limit, we can obtain an analytic expression for $g(\rho, z)$ valid for
$z\gg1/k_f^z$. We find 
\begin{equation}
\begin{split}
g(\rho,z) 
 &= g_D(\rho, z) + g_F(\rho, z)
\label{eq:g large z}\;,
\end{split}
\end{equation}
where,
\begin{equation}
g_D(\rho, z) = e^{-m_D\sqrt{\rho^2+z^2}}
\end{equation}
is the Debye screening formula which comes from the pole in the integrand in 
Eq.~\ref{eq:vmodmag} at $q = i m_D$, and 
\begin{equation}
\begin{split}
g_F(\rho,z) =& -\frac{\sqrt{\rho^2+z^2}}{\pi}\frac{\cos(2k_f^zz)}{z}
\frac{m_D^2(\pi/4)\rho}{\sqrt{(2k_f^z)^2+m_D^2(1/2)\ln(4k_f^zz)}}\\
&\times K_1(\rho\sqrt{(2k_f^z)^2+m_D^2(1/2)\ln(4k_f^zz)})
\end{split}
\end{equation}
arises from the branch cuts at $q^z=\pm 2k_f^z+i\eta$. $K$ is a modified Bessel
function of the second kind. The derivation of Eq.~$\ref{eq:g large z}$ is given in Appendix~\ref{Section:Friedel derivation}.
For $\rho=0$, we see that $g_F(0,z)$ exhibits long-range Friedel oscillations in
the $z$ direction given by
\begin{equation}
g_F(0,z) \sim -\cos(2k_f^z z)\;.
\end{equation}

Eq.~\ref{eq:g large z} is a good approximation for $z\gtrsim 1/k_f^z$. For
$\rho\rightarrow 0 $ and $z\rightarrow 0$, we expect that $g(\rho,z)$ should $\rightarrow 1$ since
the short range behavior of $V({\bf{r}})$ can not be modified by screening. For $z\rightarrow 0$
the expression Eq.~\ref{eq:g large z} breaks down because $(2k_f^z)^2+m_D^2(1/2)\ln(4k_f^zz)<0$ but in
our derivation we used the fact that $z\gtrsim 1/k_f^z$ and we can not trust Eq.~\ref{eq:g large z}
for $z<1/k_f^z$ anyway. For $\rho=0$ our derivation remains valid but Eq.~\ref{eq:g large z} can not
be used directly since $K_1(a\rho)$ goes as $1/(a\rho)$ as $\rho\rightarrow 0$. The limit of $g$ as
$\rho\rightarrow 0$, however, is finite because $a\rho K_1(a\rho)$ tends to $1$ as 
$\rho\rightarrow 0$.  Therefore we define,
\begin{equation}
\begin{split}
g(\rho=0,z)&= e^{-m_Dz}
 -\frac{\cos(2k_f^zz)}{\pi}
\frac{m_D^2(\pi/4)}{{(2k_f^z)^2+m_D^2(1/2)\ln(4k_f^zz)}} \label{eq:g rho 0}\;.
\end{split}
\end{equation}
For $z=0$ the value of $g(z)$ is well approximated by the Debye screened value
\begin{equation}
\begin{split}
g(\rho,z=0)&= e^{-m_D\rho} \label{eq:g z 0}\;.
\end{split}
\end{equation}

\begin{figure}[ht]
   \centering
   \includegraphics[width=3.5in]{rhozg_separate_eB26v2.eps} 
   \includegraphics[width=3in]{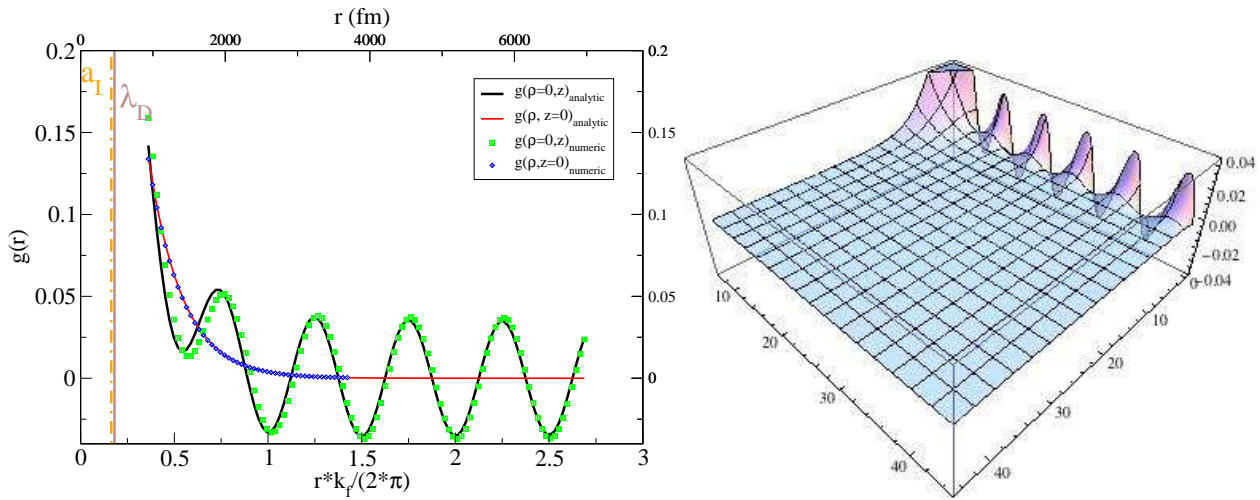} 
   \caption{(color online) Plots of $g$ as a function of $\rho$ for $z=0$, and
   as a function of $z$, for $\rho=0$ for $eB=100~eB_c$, $\mu=0.7$ MeV. $g$
   shows an exponential decay as a function of $\rho$, while along the $z$ axis
   it shows long range oscillations $\propto -\cos(2 k_f^z z)$. Also shown is
   the Debye screening length $\lambda_D$ (vertical line, brown online) and
   $a_I=(3/(4\pi n_I))^{1/3}$ (vertical dot dashed line, orange online, assuming
   $Z=26$). The solid lines correspond to the analytic approximations given in
   Eq.~\ref{eq:g rho 0} ($\rho=0$) and Eq.~\ref{eq:g z 0} ($z=0$), while the
   points correspond to a numerical computation of the Fourier transform in
   Eq.~\ref{eq:vmod}. The right panel shows the 3-d plot of $g(\rho,z)$ in the
   $\rho$, $z$ plane with $z$ and $\rho$. }  
   \label{fig:g_separate}
\end{figure}
We have calculated $g(\rho,z)$ for $eB=100 ~eB_c$ and $\mu=0.7$ MeV numerically. For these values, only
the lowest Landau level is occupied and the Fermi momentum of the lowest Landau level is $k_f^z=0.48$
MeV. So the system is mildly relativistic and the electron Fermi energy is large compared to atomic binding energy.  
The two relevant length scales for the screened potential are the Debye screening length $\lambda_D=468$ (fm), 
and the wavelength of Friedel oscillations $\lambda_F=\pi/k_f^z=eB/(2\pi
n_e)=1296$ (fm). In the left panel (Fig.~\ref{fig:g_separate})
we plot $g(\rho, z)$ along the $z$ and the $\rho$ directions separately.  This 
allows us to see the Friedel oscillations in the $z$ direction clearly, and also compare the numerical results with the approximate analytic expressions. Along the $z$ axis, Eq.~\ref{eq:g rho 0} is a good approximation to the numerical result, and in the
$x-y$ plane, the exponential Debye formula describes the result well. A 3-dimensional plot of $g(\rho,z)$ in the $\rho$, $z$ plane is shown in the right panel and depicts that the Friedel oscillations are restricted to a narrow cylinder of radius $r \simeq 1/\sqrt{eB}$ in the x-y direction. 

For large chosen field of $4.4\times10^{15}$ Gauss, which is near the upper end
of what can be found in magnetar surfaces, the amplitude of Friedel
oscillations can be very large as shown in Fig.~\ref{fig:g_separate}.
Nonetheless,  it is important to note that even small amplitude oscillations can
affect the lattice structure due to their long-range character. In our zero
temperature treatment the Friedel oscillations are undamped in the $z$ direction due
to our assumption of a sharp Fermi surface. A finite temperature will smear the
Fermi surface and this will result in exponential damping of Friedel
oscillations by the factor $\exp{(-z/\xi_T)}$, where $\xi_T=2 \pi v_{Fe}/T$ is
the thermal correlation length \cite{FetterWalecka:1971} and $v_{Fe}=k_f^z
/\mu$ is the Fermi velocity.  In neutron stars, the temperature is small and the
thermal correlation length  $\xi_T \gg a$ where $a$ is the inter-ion distance.
Consequently even for small amplitudes, Friedel oscillations can have important
effects due to their long-range nature when the screening potential from
different ions add coherently.

\section{Heat transport by longitudinal lattice phonons\label{Section:HeatTransport}}

Lattice phonons (lPhs) are space and time dependent vibrations of the ions, which can transport heat
from one region of a solid to another. Typically both longitudinal and transverse modes contribute to heat conduction. 
In this analysis we restrict ourselves to only longitudinal modes since they primarily couple to electron-hole excitations. At
 low temperatures, when Umklapp process are 
suppressed~\cite{Ziman:1972}, and the occupation numbers of lattice modes is small enough that the 
non-linear terms in the lPh lagrangian are small, the dominant scattering of lPhs is with
the electrons. We first re-derive the relation between the specific heat
conductivity of lattice phonons and the imaginary part of the electronic polarization tensor, and recall
results for scattering of lattice phonons in a free electron gas. Then we consider scattering in the presence of a magnetic field.


At low temperatures the form of the interaction term between lattice phonons and
electrons is well approximated by
\begin{equation}
{\cal{L}}_{el} = -\frac{1}{f_{el}}\int d^3 r \psi^\dagger\psi \partial_i\xi^i\;,
\end{equation}
where $1/f_{el} = (Ze^2 n_I)/(m_D^2\sqrt{m_I n_I})$, with $n_I$,
the number density of ions, $m_I$, the mass of ions, and
$m_D$, the debye screening mass~\cite{Ziman:1972}. 

The rate for phonon decay into an electron-hole ($e-h$) pair can be calculated from the
thermal width of the phonon due to absorption by electrons, and is given by the imaginary part of the 
self energy correction due to the electron loop.
\begin{equation}
\omega \Gamma (q^\mu) = -\frac{q^2}{f^2_{el}} \Im m \Pi(q^\mu) \;,~\label{eq:lPh width}
\end{equation}
where $\Gamma(q^\mu)$ is the inverse lifetime $1/\tau(q^\mu)$ of the lPh, and 
$q^\mu=(\omega, {\bf{q}})$. For an on-shell photon $\omega=c_l q$. From
Eq.~\ref{eq:lPh width}, we obtain the mean free path
$l_l(q^\mu)=\tau(q^\mu)c_l$,
\begin{equation}
\begin{split}
l_{l}(\omega, \bfq) &= -\bigl(\frac{c_l^3  f_{el}^2}{\omega \Im m \Pi(\omega, \bfq)}\bigr)\;,
\end{split}
\end{equation}
where $c_l$ is the lPh speed.

\begin{figure}[t]
   \centering
   \includegraphics[width=5in]{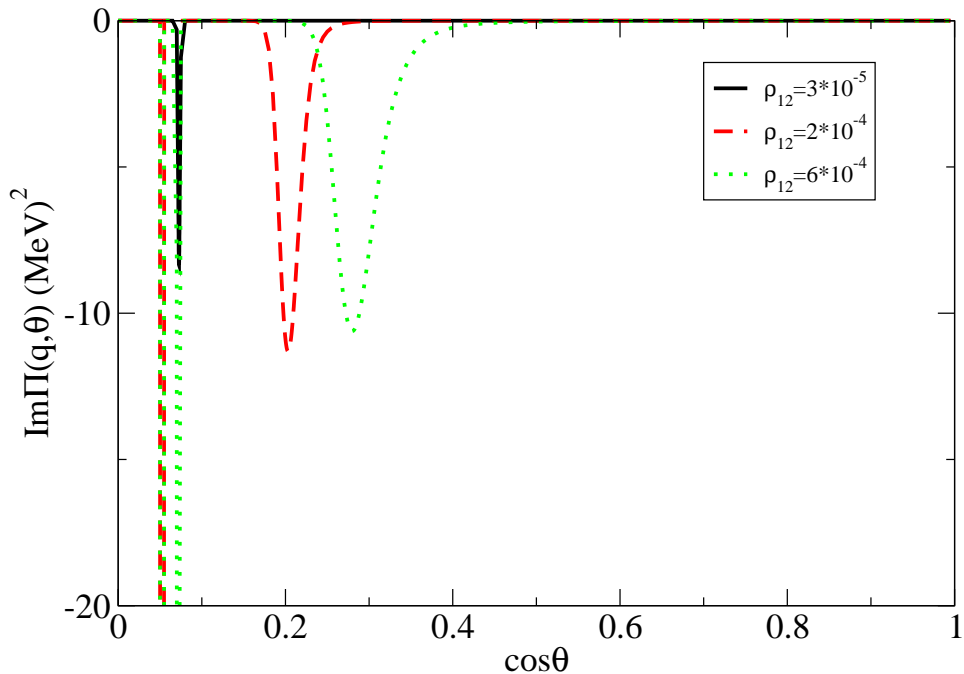} 
   \caption{(color online) A plot of $\Im
   m[\Pi(\omega=3T,q_{z}=\frac{3T}{c_s}\cos\theta,q_{\perp}=\frac{3T}{c_s}\sin\theta)]$ as a
   function of the angle $\theta$ between the magnetic field and the lPh propagation direction, for
   three different densities $\rho_{12}$(10$^{12}$gm/cc). For this plot $eB=10eB_c$, $T=5$keV, and $c_s=0.05$.
   $\cos\theta=0$ corresponds to a lPh traveling perpendicular to the magnetic field, and
   $\cos\theta=1$ corresponds to one traveling parallel to it. At
   the lowest density, only the lowest Landau level is occupied and scattering is kinematically
   allowed only at one specific angle. As more levels are occupied, more angles are kinematically allowed.}  
   \label{fig:imPivscth}
\end{figure}

The thermal conductivity contribution of the lattice phonons from kinetic theory is 
$\kappa = \frac{1}{3}C_v c_l l_{l}$. $C_v$ for lattice phonons is simply
\begin{equation}
C_v = \bigl(\frac{T}{c_l}\bigr)^3\bigl(\frac{2\pi^2}{5}\bigr)~\label{eq:C_v lPh}\;.
\end{equation}
The final expression for the specific heat conductivity for a typical phonon with $\omega=3T$ is then,
\begin{equation}
\begin{split}
\kappa = -\frac{2\pi^2}{15} T^3 \bigl(\frac{ c_l f_{el}^2}{\omega \Im m \Pi(\omega, \bfq)}\bigr) =- \bigl(\frac{c_l 2\pi^2 T^2 f_{el}^2}{45\Im m \Pi(\omega, \bfq)}\bigr)~\label{eq:kappa}\;.
\end{split}
\end{equation}

For $B=0$ the polarization function (ignoring the antiparticle contribution) of
an electron gas is well known~\cite{Jancovici:1962,FetterWalecka:1971}.
Here, for convenience, we will approximate
$\Im m \Pi(q^\mu)$ by its value at $T=0$. The corrections to this approximation
are small if $T/\mu$ is small, except at the kinematic boundaries. At $T=0$
the imaginary part of $\Pi(q^\mu)$ is, 
\begin{equation}
\Im m [\Pi(q^\mu)]  = -\frac{\mu^2\omega}{2\pi q} \theta(q v_f-|\omega|)\label{eq:ImPiB0T0}\,.
\end{equation}
For phonon decay and conductivity in Eq.~\ref{eq:kappa} we are interested in the
imaginary part for $\omega/q = c_l < v_f$, which is given by $\Im
m[\Pi(q^\mu)]= -\mu^2c_l/(2\pi)$. 



For $B\neq0$,  $\Im m[\Pi_B(q^\mu)]$ is given in Eq.~\ref{eq:ImPi}. For a representative large value of the field in a magnetar we choose  $eB=10eB_c$, and show results for the imaginary part of the
polarization tensor, and the specific heat conductivity $\kappa$.  The lattice phonons have a typical energy $\omega\sim 3T$, and we take $T=5$keV for our calculation.
The magnitude of the momentum is given by $q=\omega/c_s$, and the $z$ and the $\perp$ components are $q\cos\theta$ and $q\sin\theta$ respectively. The speed of lattice phonons depends on the
depth, but we take a representative value $c_s\sim0.05$.

\begin{figure}[t]
   \centering
   \includegraphics[width=5in]{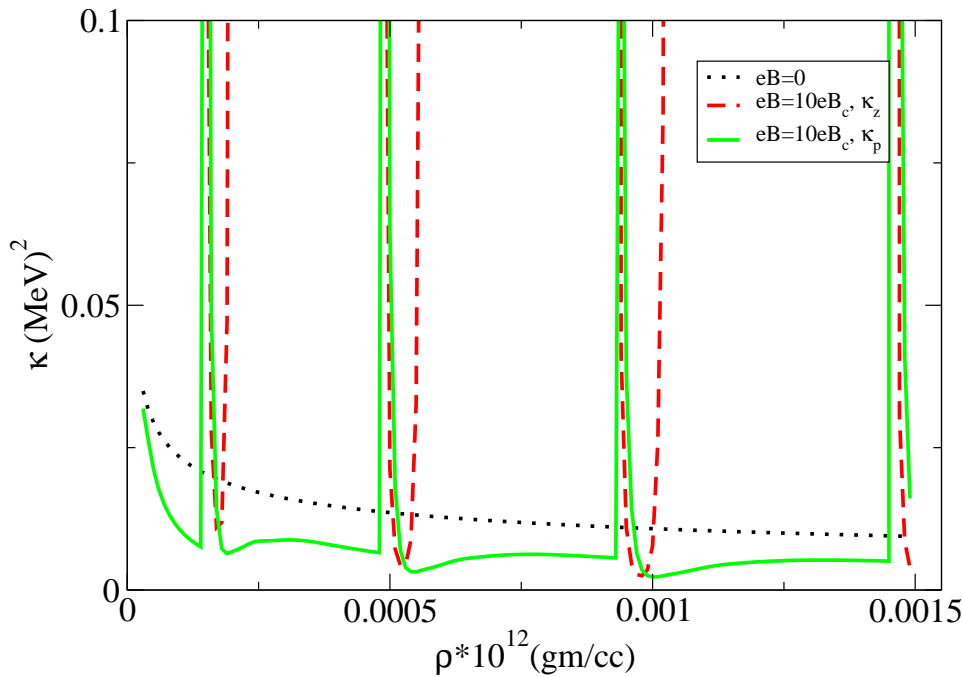} 
   \caption{(color online) Plot of $\kappa$ as a function of density. $\kappa_z$ is the specific
   heat conductivity parallel to the magnetic field, shown by the dashed (red) line. This should be
   considered as an upper bound for $\kappa_z$ because our calculation ignores Umklapp processes.
   $\kappa_\perp$ is the conductivity perpendicular the field shown by the solid line (green). The 
   dotted line (black) is the conductivity for $B=0$ at the same density.}  
   \label{fig:kappavsrho}
\end{figure}

We show the results as a function of the mass density, in commonly used units $\rho_{12}$
corresponding to $10^{12}$gm/cc. To convert the mass density, which is dominated by ions, to the
electron number density and hence the electron chemical potential, we take the atomic number of ions
to be $Z=26$ and their mass number $A=56$ (Fe).  In Fig.~\ref{fig:imPivscth}, we show $\Im m [\Pi_B(q^\mu)]$ as a function of $\cos\theta$ for a three different 
values of $\rho_{12}$. In Fig.~\ref{fig:kappavsrho} we show the specific heat conductivity parallel 
($\kappa_{z}$) and perpendicular ($\kappa_\perp$) to the magnetic field, as a function of $\rho_{12}$. From Fig.~\ref{fig:imPivscth} we see that at the lowest densities, the electrons occupy only the
lowest Landau levels, and the response is highly anisotropic, and peaked around very specific values of
$\theta$, where energy conservation and momentum conservation along the $z$ direction can be simultaneously satisfied. As the density of electrons increases, they occupy higher landau levels and overall response is obtained by summing contributions from various levels, and is non-zero for several
values of $\theta$.

To calculate  $\kappa_\perp$ and $\kappa_{z}$ in Fig.~\ref{fig:kappavsrho}, we
approximate the effective $\Im m[\Pi_B(\omega, \bfq)]$ for scattering parallel to
the magnetic field by the average between $\theta=[0,\pi/4]$, and the value
perpendicular the magnetic field by the average between $\theta=[\pi/4,\pi/2]$.
We see that the lPh conductivity is suppressed perpendicular to the magnetic
field but not parallel to it. This result is easy to see in the
ultra-relativistic limit. Energy-momentum conservation in this limit requires
$|k^z+q\cos\theta|-|k^z|=c_s q$, which can occur only at
$\cos\theta=c_s\ll\frac{1}{\sqrt{2}}$. Only at large densities where inter-level
transitions become kinematically allowed, does the net response function become
relatively isotropic. Formally, $\kappa_z$ is very large in our calculation, but
we have ignored the Umklapp processes which will be important in the $z$
direction, because electron scattering is kinematically suppressed. One can
include them by using $\kappa_z^{net}=
\kappa_z^e\kappa^U/(\kappa_z^e+\kappa^U)$, where $\kappa_z^e$ can be read from
Fig.~\ref{fig:kappavsrho}, and $\kappa^U$ is the Umklapp contribution. In the
perpendicular direction we expect electron scattering to be dominant, and it is
alright to ignore the Umklapp contribution.

 We note that averaging the response function over angles is a rough method to include the
effect of anisotropic scattering. A rigorous procedure will involve solving
the transport equation using the anisotropic collision term, that can be deduced from the
expression for $\Im m[\Pi_B(q^\mu)]$. We leave such a calculation for future work.

\section{Conclusions\label{Section:Conclusions}}

 By calculating the one-loop electron-hole polarization function in a strong
 magnetic field we have found that the screening of the ion-ion potential is
 significantly different from those obtained in earlier studies where the Debye
 approximation was assumed. We show for the first time that the Fermi surface of
 electrons in the direction parallel to the magnetic field leads to Friedel
 oscillations in the ion-ion potential. These oscillations are large and could
 fundamentally change the structure of the ionic solid at large field and
 relatively low density when only the lowest Landau levels are occupied. For
 typical magnetar field strengths of order $10^{15}$ G, matter up to densities
 of the order $10^{10}$gm$/$cc will be affected by our finding here. To evaluate
 how Friedel oscillations and modifications to the average screening length
 affects the structure and melting properties of the solid we will need to
 perform either Path Integral Monte Carlo or classical Monte Carlo simulations
 because the ion-ion interaction is highly non-perturbative. This is beyond the
 scope of this work and will be investigated separately. Here, we present
 plausible implications based on qualitative arguments to explore how our
 findings will modify the ionic structure. 

Considering the parameters used in Fig.~\ref{fig:g_separate}, $eB=100eB_c$ 
and $\mu=0.7$(MeV) and assuming $Z=26$, the average number density of ions in the system is
$n_I=3.2\times10^{-9}$ (fm)$^{-3}$. If one assumes that the ions form a regular bcc
lattice, then the separation between nearest ions is $(\sqrt{3}/2)\times
(2/n_I)^{1/3}=743$(fm).  This is comparable to the screening length
$\lambda_D$, and smaller than the wavelength for Friedel oscillations
$\lambda_F$. If significant screening in the $x-y$ plane prevents the
formation of a regular bcc structure, it may be favorable to form an
anisotropic crystal structure. One possibility is that the ions fit into the
troughs in the potential formed by superposition of the $V(z)$ separated in the
$z$ axis by $\lambda_F$. To maintain neutrality, the ions should arrange
themselves in the $x-y$ plane (in a regular or irregular structure) with an
average separation $a_\perp=(\pi n_I\lambda_F)^{-1/2} = 278$(fm). 

 If indeed the ions in the $z$ direction arrange themselves with separation
$\lambda_F$, then the following picture arises for the structure of the
ion-electron system with changing density, or as we move deeper into the neutron
star crust. A single chain of ions arranged along the $z$ axis will look like a rod.
$\lambda_F\propto eB/n_I$, and for fixed $eB$ decreases linearly with increasing
density. The transverse separation between the rods, $a_\perp\propto
(eB)^{1/2}$, on the other hand remains a constant. At the surface, it is known
that the elongated atoms form chains parallel to the $B$ field. At low densities
(near the surface of the crust), ions will form a plasma of rods with a small
charge per unit length (since $\lambda_F$ is large). These rods will interact
with each other by a two dimensional Debye screened Coulomb interaction. As we
move deeper into the star, the charge per unit length will keep on increasing,
and once $\lambda_F$ becomes comparable to $a_\perp$, the system will look more
isotropic. This picture will break down when higher Landau levels start being
occupied. It is interesting to note that for $Z=26$, $a_\perp\sim \lambda_F$
implies that $\mu^2 \gtrsim 2 eB$, meaning that the next Landau level is
occupied. Hence, the point where the screened potential becomes more isotropic
coincides with the point where the lattice structure becomes more isotropic.
Hence, in our picture, the region where the electrons occupy only the lowest
Landau level interpolates smoothly between the surface and the inner crust of
the neutron star.   

We have also showed that the heat conduction due lattice phonons can become
anisotropic because their damping rate due to electron-hole excitations is
anisotropic. It is well known that the electronic heat conductivity
perpendicular to the magnetic field is suppressed~\cite{Urpin:1980}, and this in
turn results in a temperature anisotropy at the surface of the neutron
star~\cite{Geppert:2004}. Our finding that the heat conductivity due to phonons
in the direction perpendicular to the magnetic field is also suppressed, is new.
This effect may modify the temperature anisotropy in the outer regions of the
magnetars during their early thermal evolution.

\section{Acknowledgements\label{Section:Acknowledgements}}
We acknowledge discussions with Tanmoy Bhattacharya, Joe Carlson, Stefano Gandolfi and Charles Horowitz. We thank Dima Yakovlev for notes on screening in a magnetic field. RS thanks Huaiyu Duan on discussions about the accurate implementation of Laguerre polynomials. 

\appendix

\section{Dominance of the Lowest Landau level for $eB\gg\mu^2, q^2$
\label{Section:LLL dominates}}
To see that the $n=0$ term dominates in the sum in Eq.~\ref{eq:Fb}, we look at the integrand,
\begin{equation}
\frac{W(E_0,E_n,0,n,k^z_m,k^z_n,q^{z})}{E_m-E_n}
  =\frac{(E_0E_n+k^z_mk^z_n+m_e^2)e^{-q^2l^2/2}(q^2l^2/2)^n/n!}{2E_0E_n}\frac{1}{E_n-E_m}\;,
\end{equation} 
where we have used $L_0^n(x)=1$. Now, $E_m=\sqrt{(k^z_m)^2+m_e^2}\sim\mu$. For $n=0$,
$E_n=\sqrt{(k^z+\qp)^2+m_e^2}\sim \mu$ while for $n>0$, $E_n \sim \sqrt{2neB}$. 
Therefore, for $n=0$,
\begin{equation}
\frac{W(E_0,E_n,0,n,k^z_m,k^z_n,q^{z})}{E_n-E_m}
  \sim e^{-q^2l^2/2}\frac{1}{2\mu}\;,
\end{equation} 
while for $n>0$,
\begin{equation}
\frac{W(E_0,E_n,0,n,k^z_m,k^z_n,q^{z})}{E_n-E_m}
  \sim \frac{e^{-q^2l^2/2}(q^2l^2/2)^n/n!}{2(E_n-E_m)}\sim
  \frac{e^{-q^2l^2/2}(q^2l^2/2)^n/n!}{2\sqrt{2eB}}\;.
\end{equation} 
Thus, we can conclude that the processes involving excitation to the level $n$ is suppressed
compared to the $m=0$, $n=0$ transition by a factor 
$(\mu/(\sqrt{2enB}))(q^2/(2eB))^n(1/n!)$ which is small if $eB\gg\mu^2$, $eB\gg q^2$ . 

\section{Derivation of $g(r)$ in the non-relativistic limit
\label{Section:Friedel derivation}}
To proceed with the derivation in the non-relativistic limit, we start form the expression,
\begin{equation}
\begin{split}
g(\rho,z) &= \frac{\sqrt{\rho^2+z^2}}{\pi}\int_0^\infty d\qp \qp J_0(\qp\rho)~\int_{-\infty}^\infty d\ql
\frac{\exp{(i \ql z)}}{(\ql)^2+\qp^2+F_B(\qp,\ql)} \\
 &= \frac{\sqrt{\rho^2+z^2}}{\pi}\int_0^\infty d\qp \qp J_0(\qp\rho) I(\qp,z)\;,
\end{split}
\end{equation}
where
\begin{equation}
I(\qp,z) = \int_{-\infty}^\infty d\ql \frac{\exp{(i\ql z)}}{(\ql)^2+(\qp)^2+F_B(\qp,\ql)} \;.
\end{equation}
We calculate the integral $I(\qp,z)$ by using contour integration. We assume $z>0$ and close the contour 
in the upper half of the complex plane. We have to deform the contour to go around the branch cuts 
$\pm 2k_f+i\eta$, $\eta>0$. We call these two parts of the closed contour, $C_1$ ($\qp=2k_f+i\eta$) 
and $C_2$ ($\qp=-2k_f+i\eta$). This gives
$I(\qp,z)+\int_{C_1}d\ql \frac{\exp{(i\ql z)}}{(\ql)^2+\qp^2+F_B(\qp,\ql)}+\int_{C_2}d\ql \frac{\exp{(i\ql z)}}{(\ql)^2+\qp^2+F_B(\qp,\ql)}
=2\pi i {\rm{Res}} (\frac{\exp{(i\ql z)}}{(\ql)^2+\qp^2+F_B(\qp,\ql)})
|_{\ql \ni (\ql)^2+\qp^2+F_B(\qp,\ql)=0} $. We treat these contributions one by one.

First, we look at the pole contribution which we will call the Debye part $g_D$. The pole is
where 
\begin{equation}
\begin{split}
0&=\qp^2+(\ql)^2+F_B(\qp,\ql)\\
&=\qp^2+(\ql)^2+\bigl(\frac{e}{\pi}\bigr)^2\bigl(\frac{eB}{2}\bigr)e^{-\qp^2l^2/2}
 \log\bigl(\frac{2k_f+\ql}{2k_f-\ql}\bigr)\frac{m_e}{\ql}
\end{split}
\end{equation}
If we assume that the dominant contributions comes from $\ql \ll 2k_f$, then we want 
\begin{equation}
\begin{split}
0&=\qp^2+(\ql)^2+\bigl(\frac{e}{\pi}\bigr)^2\bigl(\frac{eB}{2}\bigr)e^{-\qp^2l^2/2}
 \frac{m_e}{k_f}\\
&=\qp^2+(\ql)^2+m_D^2e^{-\qp^2l^2/2}
\end{split}
\end{equation}
This gives, $\ql=i\sqrt{\qp^2+m_D^2\exp{(-\qp^2/(2eB))}}$, where
we consider the pole in the upper half complex plane. Furthermore if the values of $\qp$ that
contribute strongly to the $d\qp$ integral are $\qp\ll \sqrt{2eB}$, we can
approximate $\exp{(-\qp^2/(2eB))}$ by $1$ and 
\begin{equation}
\begin{split}
I_D(\qp,z) &= 2\pi i {\rm{Res}} (\frac{\exp{(i\ql z)}}{(\ql)^2+\qp^2+m_D^2})\\
 & = \frac{2\pi}{2}\frac{e^{-z\sqrt{\qp^2+m_D^2}}}{\sqrt{\qp^2+m_D^2}}
\end{split}
\end{equation}
This is exactly what we obtain if we replace $F(\qp,\ql)$ by $m_D^2$ in the first place. Hence, this
integral is the same as what we obtain if we assume that there is a simple Debye like screening, and
therefore gives,
\begin{equation}
g_D(\rho,z) = e^{-m_D\sqrt{\rho^2+z^2}}
\end{equation}
This is the answer if $\qp\ll k_f,\sqrt{eB}$ dominates the $\qp$ integral, and if
$(\frac{e}{\pi})^2(\frac{eB}{2})(m_e/k_f)\ll k_f^2$.

Now we consider the integrals over the contours $C_1$ and $C_2$,
\begin{equation}
\begin{split}
I_F&(\qp,z) = \int_{C_1}d\ql \frac{e^{(i\ql z)}}{(\ql)^2+\qp^2+F_B(\qp,\ql)}+
 \int_{C_2}d\ql \frac{e^{(i\ql z)}}{(\ql)^2+\qp^2+F_B(\qp,\ql)}\\
 &=\Re e\Bigl[\int_{C_1}d\ql \frac{2e^{(i\ql z)}}{(\ql)^2+\qp^2+F_B(\qp,\ql)}\Bigr]\\
 &=\Re e\Bigl[
  \int_{\infty}^{0}d\eta\;\;2ie^{(2ik_f-\eta)z}\Bigl(\\
 &  \phantom{+}\frac{1}
     {\qp^2+(2k_f+i\eta)^2+(e/\pi)^2(eB/2)(m/(2k_f+i\eta))\exp(-\qp^2/(2eB))(1/2)(\ln((4k_f+i\eta)^2/\eta^2)-i\pi)}\\
 & -
   \frac{1}
     {\qp^2+(2k_f+i\eta)^2+(e/\pi)^2(eB/2)(m/(2k_f+i\eta))\exp(-\qp^2/(2eB))(1/2)(\ln((4k_f+i\eta)^2/\eta^2)+i\pi)}\\
 & \Bigr)\Bigr]\\
 &=\Re e\Bigl[
   \int_{0}^{\infty}
   d\eta\;\;2ie^{(2ik_f-\eta)z}\Bigl(\\
 & \phantom{+} \frac{-1}
     {\qp^2+(2k_f+i\eta)^2+m_D^2(k_f/(2k_f+i\eta))\exp(-\qp^2/(2eB))(1/2)(\ln((4k_f+i\eta)^2/\eta^2)-i\pi)}\\
 & 
   + \frac{1}
     {\qp^2+(2k_f+i\eta)^2+m_D^2(k_f/(2k_f+i\eta))\exp(-\qp^2/(2eB))(1/2)(\ln((4k_f+i\eta)^2/\eta^2)+i\pi)}
 \Bigr)\Bigr]\;.
\end{split}
\end{equation}
Because of the $\exp(-\eta z)$ in the integration over $\eta$, for $z\gg 1/k_f$ the integral is
dominated by $\eta\sim 1/z \ll k_f$. Therefore we can ignore $\eta$ whenever it is added to $k_f$. 
This simplifies the integral somewhat,
\begin{equation}
\begin{split}
I_F&(\qp,z)  \sim \Re e\Bigl[
   \int_{0}^{\infty}
   d\eta\\
 &\frac{2e^{(2ik_f-\eta) z}m_D^2\exp(-\qp^2/(2eB))(\pi/4)}
     {(\qp^2+(2k_f)^2+m_D^2\exp(-\qp^2/(2eB))(1/2)\ln(4k_f/\eta))^2+(m_D^2\exp(-\qp^2/(2eB))(\pi/4))^2}
 \Bigr]\;.
\end{split}
\end{equation}
To obtain an analytic form, we need one more simplification. We replace $\eta$ in $\ln(4k_f/\eta)$
by the value where we expect the integral to dominate, namely $\eta\sim 1/z$. The integral over
$\eta$ can then be evaluated simply, and 
\begin{equation}
I_F(\qp,z)  \sim\Bigl[
   \frac{1}{z}
 \frac{2\cos{(2k_f z)}m_D^2\exp(-\qp^2/(2eB))(\pi/4)}
     {(\qp^2+(2k_f)^2+m_D^2\exp(-\qp^2/(2eB))(1/2)\ln(4k_fz))^2+(m_D^2\exp(-\qp^2/(2eB))(\pi/4))^2}
 \Bigr]\;.
\end{equation}

The Friedel contribution to $g$ is then,
\begin{equation}
\begin{split}
g_F(\rho,z) &= -\sqrt{\rho^2+z^2}\frac{1}{\pi}
\int_{0}^{\infty} d\qp \qp J(\qp\rho) I_F(\qp,z)\\
& = -\sqrt{\rho^2+z^2}\frac{1}{\pi}\frac{2\cos(2k_fz)}{z}
\int_{0}^{\infty} d\qp \qp J(\qp\rho)\\ 
& \frac{m_D^2\exp(-\qp^2/(2eB))(\pi/4)}
     {(\qp^2+(2k_f)^2+m_D^2\exp(-\qp^2/(2eB))(1/2)\ln(4k_fz))^2+(m_D^2\exp(-\qp^2/(2eB))(\pi/4))^2}\\
\end{split}
\end{equation}
Once again assuming $\qp\ll\sqrt{eB}$ we ignore the $\exp(-\qp^2/(2eB))$ term as before. Finally, 
assuming $k_f\gg m_D$ we can drop the $(m_D^2\pi/4)$ term in the denominator, and 
\begin{equation}
\begin{split}
g_F(\rho,z)&\sim  -\sqrt{\rho^2+z^2}\frac{1}{\pi}\frac{2\cos(2k_fz)}{z}
\int_{0}^{\infty} d\qp \qp J(\qp\rho)  \frac{m_D^2(\pi/4)}
     {(\qp^2+(2k_f)^2+m_D^2(1/2)\ln(4k_fz))^2}\\
&= -\frac{\sqrt{\rho^2+z^2}}{\pi}\frac{\cos(2k_fz)}{z}
\frac{m_D^2(\pi/4)\rho}{\sqrt{(2k_f)^2+m_D^2(1/2)\ln(4k_fz)}} K_1(\rho\sqrt{(2k_f)^2+m_D^2(1/2)\ln(4k_fz)})
\label{eq:gFrhoz}\;.
\end{split}
\end{equation}

These manipulations work only if $z\gg 1/k_f$, (long distances in $z$) and $\rho\ll 1/k_f$. The
assumption $\rho\ll 1/k_f$ is not a very essential one because for $\rho\gtrsim 1/k_f$, $g_F(\rho,z)$ 
decreases very rapidly and this is captured by the expression in Eq.~\ref{eq:gFrhoz}, because of the
modified Bessel function, $K_1$.

Thus, the final answer for large $z$ is,
\begin{equation}
\begin{split}
g(\rho,z) &= g_D(\rho,z) + g_F(\rho,z)\\
 &= e^{-m_D\sqrt{\rho^2+z^2}}\\
 &-\frac{\sqrt{\rho^2+z^2}}{\pi}\frac{\cos(2k_fz)}{z}
\frac{m_D^2(\pi/4)\rho}{\sqrt{(2k_f)^2+m_D^2(1/2)\ln(4k_fz)}} K_1(\rho\sqrt{(2k_f)^2+m_D^2(1/2)\ln(4k_fz)})\;.
\end{split}
\end{equation}

\bibliography{screen} 

\end{document}